\documentclass[multphys,vecphys]{svmult}

\usepackage{makeidx}         
\usepackage{graphicx}        
\usepackage{multicol}        
\usepackage[bottom]{footmisc}
\usepackage{natbib}

\makeindex             

\begin{document}

\title*{Entropy generation in merging galaxy clusters}
\author{Michael Balogh\inst{1}\and
Ian McCarthy\inst{2}\and
Richard Bower\inst{2}\and
G. Mark Voit\inst{3}}
\institute{Department of Physics and Astronomy, University of Waterloo,  Waterloo, ON, Canada, N2L 5L4
\texttt{mbalogh@uwaterloo.ca}
\and Department of Physics, University of Durham, South Road, Durham UK, DH1 3LE
\and Department of Physics and Astronomy, MSU, East Lansing, MI 48824}
%
%
\maketitle

\section{Introduction}
This conference has seen much discussion about non-gravitational heating of the
intracluster medium, as required to reduce cooling rates and
central densities  sufficiently to explain the observed properties of
galaxy clusters. The amount of energy input required 
is often computed by comparing to the entropy
profile that would be expected from gravitational
processes alone, as determined, for example, from cosmological
simulations.  Indeed, observations of relaxed clusters show that, 
except for the innermost regions, their entropy profiles follow a
power-law profile that scales self-similarly with the cluster mass
\citep[e.g.][]{McCarthy-cooling,MFB,VKB}.

However, the origin of this default entropy profile and apparent
self-similarity is not really 
understood in detail.  Models of smooth, spherical accretion are able
to reproduce the power-law slope of the entropy gradient
\citep[e.g.][]{CMT,A+99,TN01,DSD}, but the normalization is
very sensitive to the initial gas density distribution.  In particular, if
accretion is lumpy rather than smooth (as expected in a universe
dominated by cold dark matter), insufficient entropy is
generated to explain the observations \citep{Voit03}.  If there is
such a strong dependence on initial density, it is unclear why (or if)
numerical simulations with different resolutions (and hence smoothing
scales) and implementations (e.g. Eularian or Lagrangian) are able to produce self-similar clusters.  

The source of this puzzle is illuminated by writing the equation of
hydrostatic equilibrium as
\begin{equation}
\frac{1}{\rho}\frac{d\left(\rho T\right)}{dr}=-\frac{\mu m_H}{k}\frac{GM}{r^2},
\end{equation}
which describes how the gas density and temperature ($\rho$ and $T$)
depend on distance from the cluster centre, $r$.
If the mass $M$ is dominated by dark matter, then the right hand side
is constant.  The temperature of the gas must be close to the
virial temperature $T_{\rm vir}$, again because the potential is
dominated by an external field (the dark matter).  We see, therefore,
that for any solution to this equation, an equally valid solution can
be found by scaling the density (and the corresponding boundary
conditions) by an arbitrary factor.  If we define the ``entropy'' as
\begin{equation}
K=\frac{T}{\rho^{2/3}},
\end{equation}
we see that the entropy profile of an isothermal cluster with $T\sim
T_{\rm vir}$ can also be arbitrarily
normalized and still yield a valid solution to the hydrostatic equilibrium
equation.  
Merger shocks, as
seen in both observations \citep[e.g.][]{MGBJ} and simulations
\citep[e.g.][]{RKHJ} have a very complex geometry, and entropy is
clearly not generated in a single, strong shock.
Why, then, do observed clusters show such a striking
uniformity in their entropy normalization?
 If we are to
understand how the effects of early energy injection propagate through
the mass assembly of clusters, and if
we are to implement self-consistent cooling/feedback processes in
semi-analytic models of galaxy formation, we must first understand how
entropy is generated in purely gravitational processes.  This is the
purpose of our work, recently accepted for publication
 \citep{McCarthy-mergers}, which we summarize in these proceedings.

\section{Simulations}
We have executed a number of idealized simulations of two-body cluster
mergers, using the Tree-SPH code GADGET-2 \citep{GADGET2}, as described
in detail in \citet{McCarthy-mergers}.   By default the code implements the 
entropy-conserving SPH scheme of \citet{SH03}, which ensures that the entropy of a gas particle will be conserved 
during any adiabatic process.  The simulations span a range of
mass ratios, from 10:1 to 1:1, with the mass of the primary fixed at $M_{200}=10^{15} M_\odot$.
Initial conditions and orbital parameters were chosen to match the typical conditions seen in cosmological
simulations.  Initially, the gas is assumed to be in hydrostatic
equilibrium with the dark matter, with an entropy profile similar to
that seen in observations and cosmological simulations.  

In these proceedings we discuss three interesting results from these
experiments.  These results are fairly general to all our simulations,
although here we just focus on those derived from the head-on collisions\footnote{We have also
run simulations without dark matter, which help to identify the role
played by the collisionless component.  We will not discuss those
results here, but note that our interpretation of the simulations
including dark matter often depends on what we have learned from the
gas-only runs.}.  

\subsection{Two shocks, not one}
All of our simulations show that there are {\it two} main episodes, during
which the gas entropy sharply increases.  The first
occurs when the cores of the clusters collide.  It is only then, and
not before, that there is sufficient kinetic energy to drive strong
shocks through the gas.  Therefore, the energy is not deposited
primarily at the outskirts, as assumed by spherical accretion
(``onion-skin'') models \citep{TN01,Voit03} but rather the cluster is
heated from the inside-out.  

Following this event is a more extended,
gradual increase in entropy, due primarily to the reaccretion of gas
that was driven away from the cluster by the shock wave (and
overpressurized gas at the centre) produced by the collision of the cores.  This
phase generates approximately the same amount of entropy as the initial
shock, over a longer period of time.  This phase is what we refer to as
the second shock, although in fact the 
entropy is being generated in numerous weak shocks (and perhaps
turbulent mixing) in the outer regions of the cluster\footnote{This is not
unreasonable, as the Rankine-Hugoniot equations used to determine the
post-shock conditions from the pre-shock gas and the Mach number are just a consequence of
energy and momentum conservation, and are independent of the path taken
between the initial and final state.}.

\subsection{Distributed heating}
If the gas entropy is to scale in a self-similar way following a
merger, then $K\propto T \propto M^{2/3}$.  For example, a binary merger
should result in a $\sim 60$ per cent increase in entropy.  One
remarkable result is that for all our simulations (with a range of impact
parameters and mass ratios, and even the gas-only simulations), most of
the gas satisfies this scaling after $\sim 10$ Gyr.  The exceptions are at the outer
boundary, where there is an excess of entropy generated due to
artificial boundary effects, and in the inner $\sim 10$ per cent of
gas, which is physically heated to $\sim$twice the self-similar value.

\begin{figure}
\centering
\includegraphics[height=6cm]{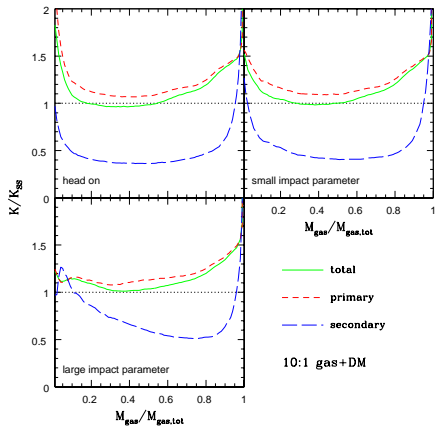}
%
%
\caption{The resulting $K(M_{\rm gas})$ distributions (entropy as a
  function of enclosed gas mass) for the 10:1 mass
  ratio mergers, normalized to $K_{200}$ (the characteristic entropy of
  the halo).  The dotted line, at $K/K_{200}=1$, is the entropy
  distribution expected if the entropy growth is self-similar.  We show the 
final entropy distributions for the primary (short-dashed),
secondary (long-dashed), and total (solid) systems for
three different orbital cases. Although most of the gas in the final
system follows the self-similar expectation, this is achieved by
overheating the primary and underheating the secondary.}
\label{fig-kmgas}       
\end{figure}

A very interesting result emerges when we simulate the merger of
clusters with unequal masses.  In this case, the final entropy profile
again agrees with the self-similar scaling law, so a 10:1 merger ends up with an entropy that is
$\sim 6$ per cent larger than that of the primary, initial halo.
One might naively have expected the gas in each component to
independently scale in this way; that is, for the gas in the small
component to be more strongly shocked than in the primary.  However,
this is not what happens.  Instead, the primary halo is {\it
  overheated}, relative to the self-similar expectation, while the
secondary is {\it underheated}, as shown in Figure~\ref{fig-kmgas}.  In other words, much of the infall
energy associated with the secondary goes into thermalising the gas in
the primary, and heating is a distributed, rather than local, process.  We find
a remarkably robust relation between the energy thermalized in both
components (primary and secondary):
\begin{equation}
\frac{E_{T,p}}{E_{T,s}}\approx \left(\frac{M_p}{M_s}\right)^{5/4}.
\end{equation}
As the mass of the primary is increased, the fraction of energy
thermalized within its gas also increases.  We do not understand why
this simple relation arises, but it seems to hold for a wide range of
mass ratios and orbital parameters.

\subsection{Energy requirements}
\begin{figure}
\centering
\includegraphics[height=6cm]{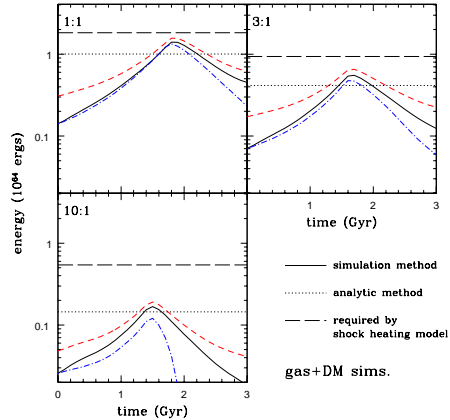}
%
%
\caption{Energy requirements in head-on mergers with three different
  mass ratios (as shown in the top-left corner of each panel).  The horizontal, dashed line shows the amount of energy
  required to produce the final entropy distribution in the
  simulations, if all the entropy is generated in a single, strong
  shock.  This is compared with the actual entropy available to be
  thermalized in the merger.  The horizontal, dotted line shows the
  result of an analytic calculation of the maximum energy available, at
  the point where the two cores collide.  The curved lines show the
  energy available as measured in the simulations, which reaches a
  maximum at the time when the cores collide.  The different line
  styles correspond to different assumptions about the interaction
  between the dark matter and gas, as described in
  \citet{McCarthy-mergers}. For the 3:1 and 10:1 simulations, there is
  not enough energy available to produce the observed entropy in a
  single shock.}
\label{fig-energy}       
\end{figure}

One would hope to be able to capture most of the relevant physics from
these simulations in a simple analytic prescription.  The first
attempts \citep[e.g.][]{Voit03} have assumed all the entropy is
generated in a single strong shock.  Voit et al. showed that, in a
simple, spherical accretion model, this fails to produce sufficient entropy if
the infalling matter is clumpy.  Our simulations show that this
conclusion still holds when we consider more realistic merger models.
In particular, for a given mass ratio, we can calculate the maximum
energy available to be thermalized, when the cores collide, as shown in
Figure~\ref{fig-energy}.  If we then
assume that all of this energy is thermalized in a single, strong
shock, we are unable to produce enough entropy to explain the results
of 3:1 or 10:1 mergers.  

This is a puzzling result, as our naive expectation had been that a
single, strong shock thermalizing all the available energy would yield
the maximum entropy.  This is true, in that thermalizing the energy in
$N$ weaker shocks does not produce as much entropy, if the system does
not evolve between shocks.
Where does the real system find the extra energy?  The answer appears
to be that the gas density actually decreases significantly between the
first and second shock events.  In \citet{McCarthy-mergers}, we show
analytically that if the density drops 20-30\% below its {\it
  pre-merger} value, then the second
shock can generate enough entropy for the final cluster to attain its
self-similar structure.  From the simulations we directly measure that
this is indeed what happens: the first shock actually drives gas
outward, so that the density in the outer regions ends up lower than it
was initially, by 20-30\%.

\section{Conclusions}
Gravitational shock heating of clusters does not appear to be as simple
a process as once envisaged.  It is crucial that we understand this
mechanism, if we are to improve semi-analytic models of galaxy formation
and to understand the heating requirements of real clusters.  Our
simulations have shown that self-similarity is indeed achieved during
cluster mergers, but that this does not happen in a single accretion
shock, because there is insufficient energy available.
Instead, entropy is generated in two major ``shocks'', that heat the gas from the
inside-out, in a way that distributes most of the energy within the
more massive clump.  

Our ultimate goal is to construct an analytic model of cluster growth
that self-consistently tracks the entropy of the gas as it is shocked
or non-gravitationally heated.  There is still much work to be done
before we achieve this.  
Our next steps are to test our analytic, two-shock model against
idealized simulations with perturbed initial conditions
(e.g. preheating).  

%
%
%

\vskip 1cm
We would like to thank our collaborators F. Pearce, T. Theuns,
A. Babul, C. Lacey and C. Frenk, who are co-authors on this work as
submitted in \citet{McCarthy-mergers}.
%
%
%
\bibliographystyle{apj}
\bibliography{ms}

\begin{thebibliography}{13}
\expandafter\ifx\csname natexlab\endcsname\relax\def\natexlab#1{#1}\fi

\bibitem[{{Abadi} {et~al.}(2000){Abadi}, {Bower}, \& {Navarro}}]{A+99}
{Abadi}, M.~G., {Bower}, R.~G., \& {Navarro}, J.~F. 2000, MNRAS, 314, 759

\bibitem[{{Cavaliere} {et~al.}(1998){Cavaliere}, {Menci}, \& {Tozzi}}]{CMT}
{Cavaliere}, A., {Menci}, N., \& {Tozzi}, P. 1998, ApJ, 501, 493

\bibitem[{{Dos Santos} \& {Dor{\'e}}(2002)}]{DSD}
{Dos Santos}, S. \& {Dor{\'e}}, O. 2002, A\&A, 383, 450

\bibitem[{{Markevitch} {et~al.}(2005){Markevitch}, {Govoni}, {Brunetti}, \&
  {Jerius}}]{MGBJ}
{Markevitch}, M., {Govoni}, F., {Brunetti}, G., \& {Jerius}, D. 2005, ApJ, 627,
  733

\bibitem[{{McCarthy} {et~al.}(2005){McCarthy}, {Fardal}, \& {Babul}}]{MFB}
{McCarthy}, I., {Fardal}, M., \& {Babul}, A. 2005, ApJ, submitted,
  astro-ph/0501137

\bibitem[{{McCarthy} {et~al.}(2007)}]{McCarthy-mergers}
{McCarthy}, I. {et~al.} 2007, MNRAS, in press, astro-ph/0701335

\bibitem[{{McCarthy} {et~al.}(2004){McCarthy}, {Balogh}, {Babul}, {Poole}, \&
  {Horner}}]{McCarthy-cooling}
{McCarthy}, I.~G., {Balogh}, M.~L., {Babul}, A., {Poole}, G.~B., \& {Horner},
  D.~J. 2004, ApJ, 613, 811

\bibitem[{{Ryu} {et~al.}(2003){Ryu}, {Kang}, {Hallman}, \& {Jones}}]{RKHJ}
{Ryu}, D., {Kang}, H., {Hallman}, E., \& {Jones}, T.~W. 2003, ApJ, 593, 599

\bibitem[{{Springel}(2005)}]{GADGET2}
{Springel}, V. 2005, MNRAS, 364, 1105

\bibitem[{{Springel} \& {Hernquist}(2003)}]{SH03}
{Springel}, V. \& {Hernquist}, L. 2003, MNRAS, 339, 289

\bibitem[{{Tozzi} \& {Norman}(2001)}]{TN01}
{Tozzi}, P. \& {Norman}, C. 2001, ApJ, 546, 63

\bibitem[{{Voit} {et~al.}(2003){Voit}, {Balogh}, {Bower}, {Lacey}, \&
  {Bryan}}]{Voit03}
{Voit}, G.~M., {Balogh}, M.~L., {Bower}, R.~G., {Lacey}, C.~G., \& {Bryan},
  G.~L. 2003, ApJ, 593, 272

\bibitem[{{Voit} {et~al.}(2005){Voit}, {Kay}, \& {Bryan}}]{VKB}
{Voit}, G.~M., {Kay}, S.~T., \& {Bryan}, G.~L. 2005, MNRAS, 364, 909

\end{thebibliography}
%

\printindex
\end{document}